\documentclass[%
reprint, 
amsmath,amssymb,
aps, physrev,
floatfix,
]{revtex4-2}

\usepackage{graphicx} 
\usepackage{dcolumn} 
\usepackage{bm} 
\usepackage{natbib}
\usepackage{xcolor}

\begin{document}

\preprint{APS/123-QED}

\title{\textbf{ Logistic dynamics of small populations \\ with demographic stochasticity} 
}

\author{Lucas M. Brugevin}
\author{Dami\'an H. Zanette}
\altaffiliation[Also at ]{Consejo Nacional de Investigaciones Cient\'{\i}ficas y T\'ecnicas, Argentina}
 \email{Contact author: damian.zanette@ib.edu.ar}
\affiliation{
 Centro At\'omico Bariloche and Instituto Balseiro, \\ Comisi\'on Nacional de Energ\'{\i}a At\'omica and Universidad Nacional de Cuyo, \\ 8400 San Carlos de Bariloche, R\'{\i}o Negro, Argentina
}%

\date{\today}

\begin{abstract}
We study an ecology-inspired model for a population of bounded size, whose dynamics is 
governed by random birth, death, and immigration events. Stochastic fluctuations in the number of individuals give rise to a succession of alternating active and vacant periods, where the population is respectively extant and extinct. Using both analytical and numerical techniques, we characterize the statistics of the two kinds of period, quantifying their duration and frequency, and the typical  population sizes in active periods. In sharp contrast to the deterministic mean-field behavior, governed by logistic dynamics, active periods may exhibit pronounced bimodality: either short durations with very small populations, or much longer durations with population sizes close to the maximum. We also investigate how these results change when the population evolves on random networks of three classes: Erd\H{o}s-R\'enyi, regular, and geographic. The main effect of the network structure is to induce population clustering, with individuals aggregated into localized groups. This, in turn, limits population growth and increases the frequency of vacant periods. 
\end{abstract}

\maketitle

%\tableofcontents

\section{Introduction}

The study of small populations of living beings is of crucial importance in several ecological scenarios, encompassing endangered species, fragmented communities, and ecosystems in extreme environmental conditions \cite{Lande,Imeson,Jeppsson}. Unlike large populations, which can be modeled with mean-field formulations that average over many individuals, small populations are dominated by random fluctuations and variability at the individual level. In such cases, demographic stochasticity, environmental noise, and chance events such as aleatory extinction or successful colonization play a leading role \cite{Morris,Lande,Nature}. Mean-field approaches \cite{Murray} smooth out these fluctuations, predicting continuous and deterministic trajectories that may not capture the fragility of small populations. For example, a mean-field model might suggest that a population will stabilize at a given equilibrium size, while a small population could collapse to extinction due to a run of bad years or the accidental loss of a few reproductive individuals. Such effects may have a direct impact on the evaluation of ecosystem vulnerability, extinction risks, critical thresholds, and the effectiveness of conservation strategies.

A similar argumentation applies to socioeconomic phenomena in small human populations. In such groups, individual decisions and random interactions can entail disproportionate consequences, leading to outcomes such as sudden consensus, rapid polarization, or collapse of cooperation \cite{PRE}. Both in ecological and socioeconomic mathematical modeling, the approach to this kind of effect calls for a formulation based on stochastic processes, tailored to capture discrete, probabilistic events.

From a mathematical-physics standpoint, one of the earliest systematic approaches to incorporating stochasticity into population dynamics was van Kampen's $\Omega$-expansion, originally developed in the context of chemical kinetics \cite{vK81}. The method relies on an expansion of the population density in inverse powers of the square root of the system size, $\Omega$. The leading-order terms reproduce the deterministic mean-field equations, while higher-order contributions yield Gaussian and non-Gaussian corrections. More recent variations of this formulation \cite{SSG} are currently being applied in mathematical ecology (see, for instance, Refs. \citenum{vKa} and \citenum{vKb}).
 
The introduction of the perturbative parameter $\Omega^{-1/2}$ in Van Kampen's expansion, however, results from an ad-hoc assumption on the dependence of fluctuations on the system size. More ab-initio approaches can be formulated starting from intrinsically stochastic models, in particular, from computational simulation algorithms with random dynamics, as originally proposed by Gillespie for chemical reactions \cite{Gillespie}. In these models, the population size --or its maximum value-- can be introduced as one of the control parameters. Moreover, at least to some extent, they can be analyzed using standard tools from the theory of stochastic processes. This methodological framework has given rise to a broad class of both analytical and computational models, whose variants have been adapted to describe phenomena of ecological and evolutionary relevance such as predator-prey interactions \cite{prpr,prpr1,prpr2}, epidemic spread \cite{epid2,epid1,epid}, and game-like dynamics \cite{Araujo}, among many others \cite{others1,others2,others21,others3,others31,others32,others4}. 

In this contribution, we examine the basic ecological setting of a single species whose population dynamics are driven by birth-death stochastic competition, in an environment that limits the population size to a maximum. In the mean-field limit, this system exhibits logistic dynamics \cite{Murray}, with a carrying capacity which derives from the maximum population size \cite{others1}. In contrast with the mean-field behavior, however, stochastic fluctuations inevitably lead the finite-size population to extinction. To prevent the system from reaching this absorbing stationary state, we incorporate random immigration events that occasionally introduce new individuals into the population. Immigration is assumed to occur far less frequently than birth and death, yet it still produces nontrivial long-time statistics. Stochastic logistic-like dynamics has been examined in recent work \cite{log1,log2,log3,log4}, often with emphasis on population persistence versus extinction.   

Here we present a detailed analytical and numerical study of the long-time statistical behavior of this class of systems, with particular emphasis on its dependence on the maximum population size and on the relative rates of birth and death. We consider both unstructured systems and spatially distributed populations on three classes of random networks. Whenever possible, we provide a comparison between stochastic and mean-field results.

\section{Model and results for  unstructured populations} \label{II}

Our dynamical model is based on a variant of the Gillespie-inspired scheme introduced and discussed in Ref.~\citenum{others1}. We consider a population of individuals occupying a set of $N$ sites. In the version considered in this section, this set lacks any particular structure, in the sense that the spatial arrangement of sites plays no role in the population dynamics. Each site may be empty or occupied by at most one individual; therefore, the total population size cannot exceed $N$ individuals. Thus, $N$ can be interpreted as the carrying capacity of the environment supporting the population. 

The dynamics proceed in discrete time steps, encompassing events of individual birth, death, and immigration from outside the system. At each step:
\begin{itemize}
    \item With probability $u$, a site is chosen at random from the whole system.
    \begin{itemize}
        \item Death: If the chosen site is occupied, it becomes empty with probability $p_\mu$.
        \item Immigration: If the site is empty, it becomes occupied with probability $p_\alpha$.
    \end{itemize}
    \item With the complementary probability, $1-u$, an ordered pair of sites is chosen at random from the whole system.  
    \begin{itemize}
        \item Birth: If the first site is occupied and the second is empty, the latter becomes occupied with probability $p_\nu$.
    \end{itemize}
\end{itemize}

Splitting the process into the two sets of events with complementary probabilities $u$ and $1-u$ is aimed at organizing the decision tree needed to apply the above dynamical rules. As shown later, these probabilities become ultimately combined with $p_\mu$, $p_\alpha$ and $p_\nu$, so that their specific values can be chosen with ample freedom. In the numerical results presented throughout this work, we always take $u=1-u=0.5$. 

Note that, unlike the standard deterministic formulation of logistic dynamics, the stochastic process considered here allows for population growth through immigration. This inclusion is justified because, given a finite per-individual death probability at each time step (provided $u\neq 0$ and $p_\mu\neq 0$), the only stationary state in the absence of immigration is total extinction: without incoming individuals, the population inevitably vanishes at sufficiently long times. Immigration counteracts this effect by enabling the population to reappear after extinction. Nevertheless, since our main interest lies in the interplay between birth and death in a limited environment --that is, the traditional logistic setting-- we restrict the analysis to the regime of relatively infrequent immigration: $p_\alpha \ll p_\mu , p_\nu$.   

\begin{figure}[ht]
\includegraphics[width=\columnwidth]{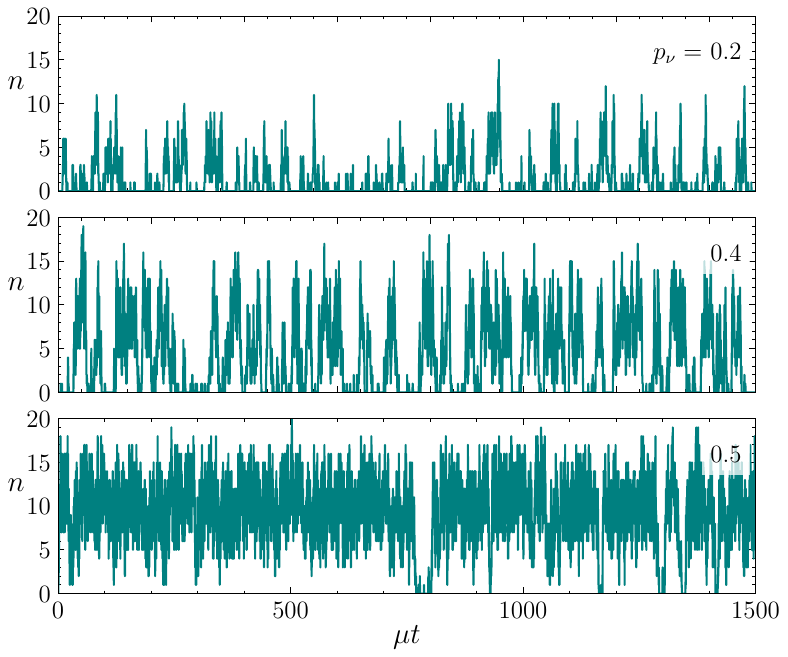}
\caption{\label{fig01} Numerical results from realizations of the model for the population size $n$ as a function of the rescaled time $\mu t$ (see main text), in a system with carrying capacity $N=20$, immigration probability $p_\alpha=0.002$, death probability $p_\mu=0.2$, and three values of the birth probability $p_\nu$. 
} 
\end{figure}

Figure \ref{fig01} shows the number of occupied sites $n$ as a function of time, along three realizations of the model with $N=20$, $p_\mu = 0.2$ $p_\alpha=0.002$, and three values of $p_\nu$. The horizontal axis is scaled in such a way that the unit of time coincides with the average lifespan of an individual, given by
\begin{equation} \label{life}
    \mu^{-1} = \frac{N\delta t}{up_\mu},
\end{equation}
where $\delta t$ is the duration assigned to each time step. In all cases, the evolution consists of a sequence of {\em active} periods, where $n(t)>0$, punctuated by {\em vacant} periods, with $n(t)=0$. As the birth probability $p_\nu$ grows, active periods become longer and vacant periods occur less frequently. Concurrently, the typical values attained by $n(t)$ increase. 

\begin{figure}[ht]
\includegraphics[width=\columnwidth]{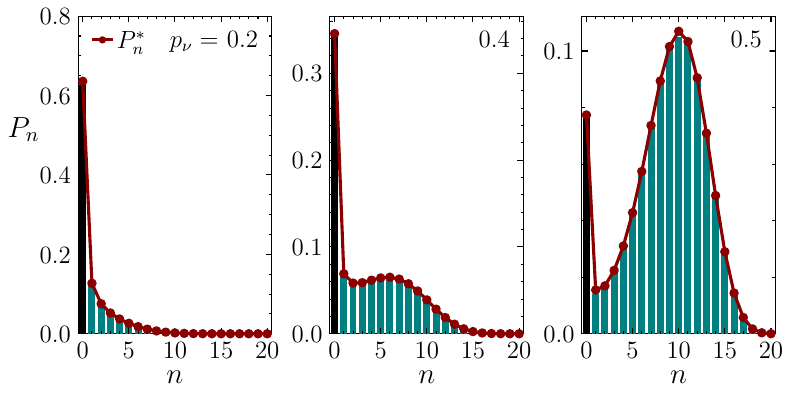}
\caption{\label{fig02} Normalized histograms showing the frequency $P_n$ of each population size $n$ along numerical realizations of the model for the same parameters as in Fig.~\ref{fig01}. The detachment of $P_0$ from the rest of the histogram is highlighted by a different color for the respective column.  Dots joined by lines correspond to the values $P_n^*$ predicted by the stationary solution to the master equation (\ref{ME}).} 
\end{figure}

The normalized histograms of Fig.~\ref{fig02} show the frequencies --namely, the probabilities-- $P_n$ of each value of $n$ measured along very long realizations of the model, with the same parameters as in Fig.~\ref{fig01}. It is apparent that, in the range of variation of $p_\nu$ considered here, a qualitative change in the statistics of $n(t)$ occurs, with the appearance of a well-developed peak of probability for $n>0$ as $p_\nu$ increases. At the same time, the value of $P_0$ clearly lies outside the general trend of $P_n$ for positive $n$, producing a distinct peak on the left side of the histograms. Note that $P_0$ is simply the fraction of time the system spends in vacant periods.

It turns out that the behavior of $P_n$ can be accurately described by the stationary state of the master equation for the stochastic process that defines our model. The form of the equation and its long-time solution are discussed in the following.

\subsection{Master-equation approach} \label{IIA}

Taking into account the Markovian nature of the stochastic process defined above, and the fact that at most one individual is added or removed from the population at each time step, we can immediately obtain the corresponding master equation for the probabilities $P_n(t)$, as \cite{others1,others11}
\begin{eqnarray}
    \dot P_n = &&f_+ (n-1) P_{n-1} \bar \delta_{n0}+f_-(n+1)  P_{n+1} \bar \delta_{nN}\nonumber\\ &&- [f_+(n)+f_-(n)] P_n, \label{ME}
\end{eqnarray}
for $n=0,1,\dots, N$, with
\begin{equation} \label{fs}
    f_+(n)= \alpha (N-n) +\nu n \frac{N-n}{N-1}, \ \ \ \ \ f_-(n)= \mu n ,
\end{equation}
and $\bar \delta_{nm}=1-\delta_{nm}$ the complement of the Kronecker delta. Moreover,
\begin{equation} \label{alphanu}
    \alpha= \frac{up_\alpha}{N\delta t}, \ \ \ \ \ \nu=\frac{(1-u)p_\nu}{N \delta t},
\end{equation}
and $\mu$ is defined as in Eq.~(\ref{life}). The master equation (\ref{ME}) is written in the continuous-time limit, where the duration $\delta t$ assigned to each time step tends to zero.

From Eqs.~(\ref{ME}), it is possible to obtain an equation of motion for the mean value of the number of individuals, $\langle n\rangle=\sum_{n=0}^N n P_n$, which, due to the nonlinearity of $f_+(n)$, turns out to depend also on the second-order moment $\langle n^2 \rangle$. Approximating $\langle n^2 \rangle \approx \langle n\rangle^2$, as expected to be valid in the limit of large carrying capacity \cite{others1}, $N\to \infty$, we obtain a mean-field equation for the density $\rho =\langle n\rangle/N$:
\begin{equation}
    \dot \rho = \nu \rho (1-\rho)-\mu \rho+ \alpha (1-\rho) .
\end{equation}
For $\alpha=0$, the mean-field equation is purely logistic, with a stable fixed point at
\begin{equation}
    \rho^*_{\rm MF} = 
    \begin{cases}
        0,  &\mbox{for $\nu \le \mu$}, \\
        1-\mu/\nu, &\mbox{for $\nu > \mu$}.
    \end{cases}
\end{equation}
At the critical point $\nu=\mu$, it exhibits the usual transcritical bifurcation of logistic systems \cite{Drazin}. For $\alpha\neq 0$, on the other hand, the bifurcation disappears, and the stable equilibrium becomes
\begin{equation} \label{rhoMF}
\rho^*_{\rm MF} = \frac{(\nu-\mu-\alpha)+\sqrt{(\nu-\mu-\alpha)^2+4\nu\alpha}}{2\nu}.
\end{equation}
Later on, we compare these results for the mean-field dynamics to those obtained for the stochastic model with finite $N$. 

In principle, the stationary solution of the master equation (\ref{ME}), $\dot P_n=0 \ \forall n$, can be solved recursively, starting from the equation for $n=0$, $\mu P_1^* -\alpha N P_0^*$, and assuming that $P_0^*$ is known. This recursive procedure makes it possible to immediately show, first, that the only solution when $\alpha=0$ is $P_0^*=1$ and $P_n^*=0 \ \forall n\ge 1$. Indeed, as advanced above, the only stationary solution in the absence of immigration is the disappearance of the whole population. Moreover, in the limit $\alpha \to 0$ --concretely, for $\alpha \ll \nu /(N-1)$-- an approximate stationary solution to the master equation for $n\ge 1$ is
\begin{equation} \label{Pnapp}
    P_n^* = \frac{N!}{(N-n)!} \frac{\alpha  \nu^{n-1} P_0^*}{n \mu^n (N-1)^{n-1}}.
\end{equation}
The value of $P_0^*$ can subsequently be found by requiring normalization of the stationary probability, $\sum_{n=0}^N P_n^*=1$. This latter problem, however, does not seem to have an analytic solution.

For finite values of $\alpha$, the stationary solution of Eqs.~(\ref{ME}) must also be found numerically. Being a linear algebraic problem equivalent to finding the kernel of an $N\times N$ matrix, it can be efficiently solved using a variety of numerical algorithms, even for large $N$. Note that the solution is not independently determined by the three parameters $\mu$, $\alpha$, and $\nu$, but by two suitable combinations of them, for instance, $\alpha/\mu$ and $\nu/\mu$. This specific choice amounts to fixing the time unit equal to the lifespan $\mu^{-1}$, as already done in Fig.~\ref{fig01}. Dots joined by lines in Fig.~\ref{fig02} stand for the values of $P_n^*$ obtained from the numerical stationary solution to the master equation (\ref{ME}), for the same parameters as the corresponding histograms. The excellent agreement with the results obtained from the numerical realizations of the model is apparent. This coincidence makes it possible to characterize the statistics of the population size $n$ along realizations directly using the stationary solution of the master equation, instead of the realizations themselves, which are computationally much more demanding. In the following, we adopt this approach to analyze the relative duration of active and vacant periods and the long-time statistical properties of $n$.  

\subsection{Vacant time and average activity} \label{IIB}

As a first characterization of the dynamics of our model, we consider the fraction of time that the system spends in vacant periods, namely, $P^*_0$. In the results shown below, we fix $\alpha/\mu=0.01$, and study the dependence of $P^*_0$ on the birth-to-death frequency ratio $\nu/\mu$ and the carrying capacity $N$. 

\begin{figure}[ht]
\includegraphics[width=\columnwidth]{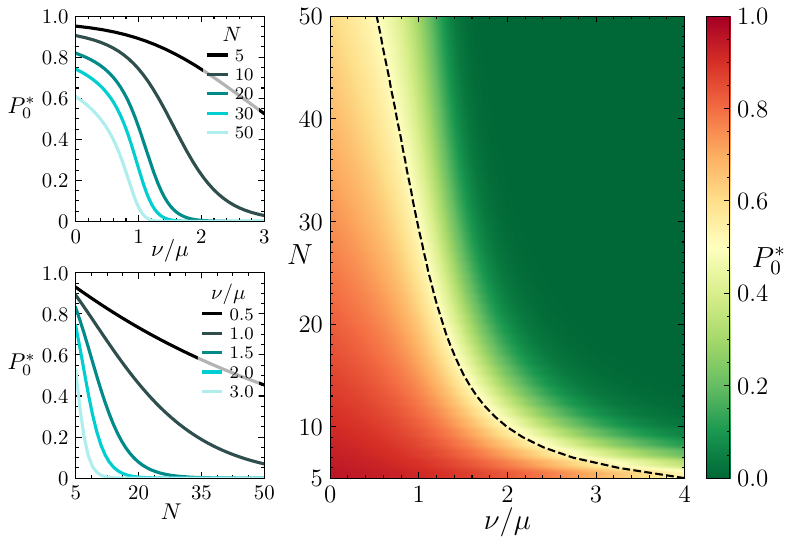}
\caption{\label{fig03} Left column: Fraction of time spend by the population in vacant periods, given by the stationary solution to the master equation (\ref{ME}), $P_0^*$, as a function of $\nu/\mu$ for five values of $N$ (upper panel), and as a function of $N$ for five values of $\nu/\mu$ (lower panel). Right: Color map of $P_0^*$ on the plane $(\nu/\mu,N)$. The dashed curve stands for the manifold where $P_0^*=0.5$, which separates the regions where the total duration of vacant periods prevails over that of active periods, to the left of the curve, and vice versa to the right.} 
\end{figure}

The upper-left panel of Fig.~\ref{fig03} shows $P_0^*$ as a function of $\nu/\mu$ for five values of $N$. As expected, the fraction of time in vacant periods decreases as the birth frequency grows, from a finite level for $\nu/\mu \approx 0$, to zero for sufficiently large values of $\nu/\mu$. 
%For $\nu/\mu \to 0$ and small $\alpha$, $P_0^*$ can be approximated using Eq.~(\ref{Pnapp}), namely,  $P_0^*(\nu/\mu \to 0) \approx (1+N\alpha/\mu)^{-1}$.
As $N$ increases, the interval where $P_0^*$ drops becomes narrower and closer to $\nu/\mu=1$. According to the mean-field results discussed in Section \ref{IIA}, in fact, for $N\to \infty$ and $\alpha\to 0$, $P_0^*$ should display a sharp discontinuity, dropping to zero for $\nu/\mu>1$. The lower-left panel of Fig.~\ref{fig03}, meanwhile, shows $P_0^*$ as a function of $N$ for five values of $\nu/\mu$. We see that $P_0^*$ decreases monotonically to zero as $N$ grows, with a steeper descent for larger birth frequencies. Now, however, the curves exhibit no inflection points, suggesting that a ``critical'' value of $N$ does not exist. 

The rightmost panel in Fig.~\ref{fig03} integrates the above information in a color map for $P_0^*$ on the plane spanned by $\nu/\mu$ and $N$. The dashed curve indicates the manifold where $P_0^*=0.5$. To the left of this curve, the population spends longer times in vacant periods than in active periods, and vice versa. In other words, both larger birth frequencies and carrying capacities promote the persistence of the population. 

To characterize the population during active periods, we use the rescaled density
\begin{equation} \label{rhoact}
    \rho^*_{\rm  act} = \frac{1}{N(1-P_0^*)} \sum_{n=1}^N nP_n^*,
\end{equation}
which disregards the contribution of $n=0$, and the corresponding standard deviation,
\begin{equation} \label{sigact}
    \sigma_{\rm act}^* = \left[ \frac{1}{N^2(1-P_0^*)} \sum_{n=1}^N n^2P_n^*-\left( \rho^*_{\rm  act}\right)^2 \right]^{1/2}.
\end{equation}
These two quantities correspond, respectively, to the mean density and its standard deviation computed over the course of the evolution, excluding vacant periods. We also introduce their coefficient of variation $V_{\rm act}^*= \sigma_{\rm act}^*/\rho^*_{\rm  act} $.

\begin{figure}[ht]
\includegraphics[width=.75\columnwidth]{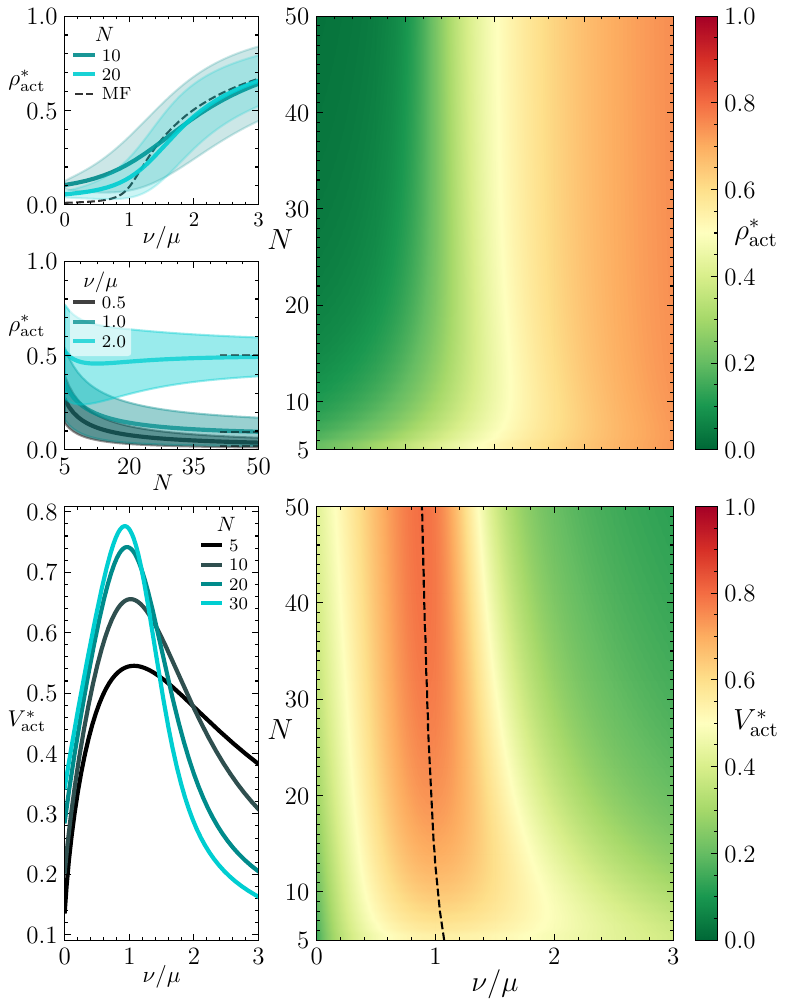}
\caption{\label{fig04}Upper half: Average density during active periods, $\rho_{\rm act}^*$, defined as in Eq.~(\ref{rhoact}), as a function of $\nu/\mu$, for two values of $N$ (upper-left), and as a function of $N$ for three values of $\nu/\mu$ (lower left), with $\alpha/\mu=0.01$. Translucent bands represent the corresponding standard deviation $\sigma^*_{\rm act}$, Eq.~(\ref{sigact}). In the former plot, the dashed curve shows the mean-field result of Eq.~(\ref{rhoMF}). In the latter, the horizontal dashed segments stand for the mean-field result for each value of $\nu/\mu$. The color map to the right shows $\rho_{\rm act}^*$ on the plane spanned by $\nu/\mu$ and $N$. Lower half: Coefficient of variation of the density during active periods, $V_{\rm act}^*= \sigma_{\rm act}^*/\rho^*_{\rm  act} $, as a function of $\nu/\mu$  for several values of $N$ (left), and as a color map on the plane spanned by $\nu/\mu$ and $N$ (right). In the latter, the dashed curve indicates the position of the maximum of $V^*_{\rm act}$.} 
\end{figure}

Full curves in the upper-left panel of Fig.~\ref{fig04} show the average density during active periods $\rho^*_{\rm act}$ as a function of $\nu/\mu$ for two values of $N$.  Translucent bands represent the corresponding standard deviation $\sigma^*_{\rm act}$. Their main mutual difference occurs for $\nu/\mu \lesssim 1$, where $\rho^*_{\rm act}$ decreases for large $N$. This trend is confirmed by the mean-field result ($N\to \infty$), Eq.~(\ref{rhoMF}), shown by the dashed curve. The middle-left panel of the same figure shows $\rho^*_{\rm act}$ as a function of $N$ for three values of $\nu/\mu$. The horizontal dashed segments to the right are the expected mean-field values. We see that the largest variation of $\rho_{\rm act}^*$ with $N$ is concentrated in relatively small values of the carrying capacity, $N\lesssim 10$. This is also apparent in the color map to the right, where the rather sharp dependence on $\nu/\mu$ contrasts with a virtual independence of $N$, except in the lower end of the vertical axis.   

As for the fluctuations of the density in active periods, the lower-left plot of Fig.~\ref{fig04} shows its coefficient of variation $V_{\rm act}^*$ as a function of the birth rate $\nu/\mu$ for five values of $N$. Its main feature is a peak at $\nu/\mu \approx 1$, which becomes sharper as $N$ grows and the mean-field limit is approached. This maximum in the relative fluctuations of the density is consistent with the critical phenomenon that characterizes the mean-field dynamics when $\alpha=0$, namely, the transcritical bifurcation at $\nu/\mu=1$, separating the phases of population extinction and persistence. In the lower-right panel, $V_{\rm act}^*$ is plotted as a color map, with the dashed line indicating the position of the maximum. 

\subsection{Statistics of vacant and active periods} \label{IIC}

The quantities analyzed in Section \ref{IIB} account for average features of the evolution along extended time intervals. We have shown that they are well described by the stationary probabilities of population sizes, as given by the time-independent solution of the master equation (\ref{ME}). A more detailed characterization of the dynamics requires examining the succession of vacant and active periods and recording their individual durations. Within each active period, moreover, the population size attained can also be studied.

\begin{figure}[ht]
\includegraphics[width=.8\columnwidth]{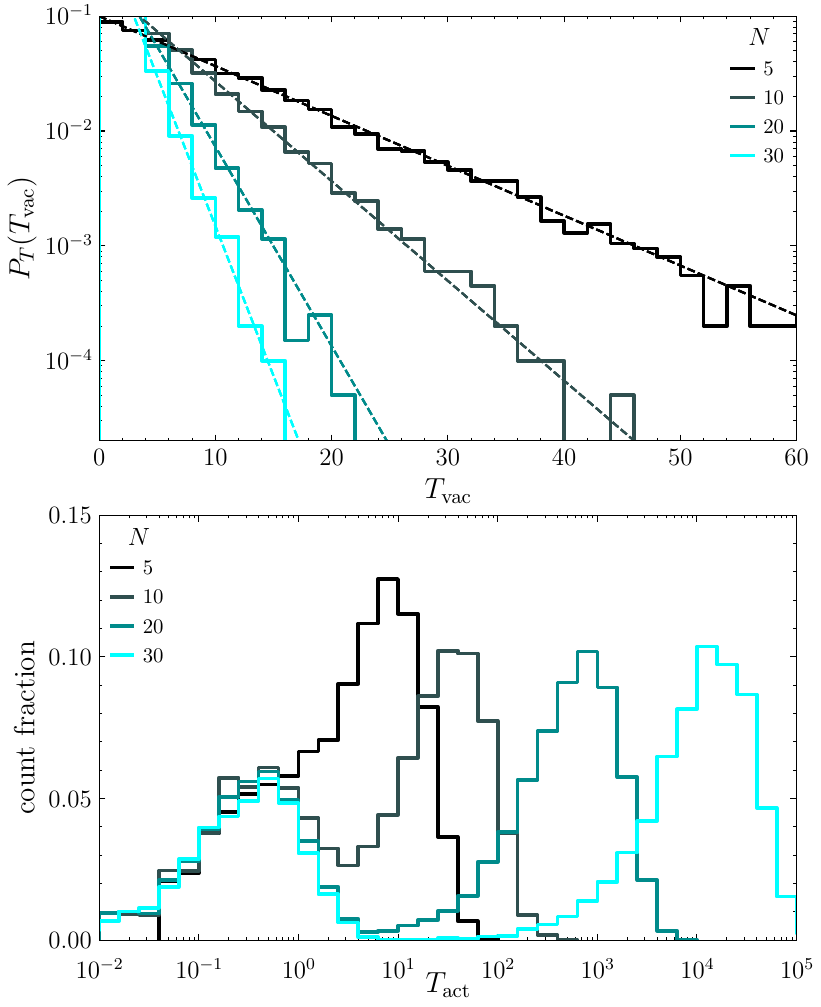}
\caption{\label{fig05}Upper panel: Normalized histograms for the duration of vacant periods, $T_{\rm vac}$, constructed from $10^4$ realizations for each of four values of $N$, with the same parameters as in Fig.~\ref{fig01}. Dashed straight lines stand for the analytical prediction of Eq.~(\ref{Pvac}). Lower panel: Histograms for the active times, $T_{\rm act}$, with the same parameters as in the upper panel. For a clearer interpretation, the vertical axis shows the ratio between the number of  counts in each column and the total number of active periods in each histogram. Note that, due to the logarithmic scale in the horizontal axis, columns actually have different widths in the variable $T_{\rm act}$. 
} 
\end{figure}

The probability $p(s_{\rm vac})$ that a given vacant period lasts  $s_{\rm vac}$ time steps is immediately found by noting that, once the population size has reached the value $n=0$, the only process that can terminate the vacant period is immigration. Recalling that, when $n=0$, the probability of an immigration event is $up_\alpha$ per time step, we get the geometric distribution
\begin{equation}
    P_s(s_{\rm vac}) = up_\alpha (1-up_\alpha)^{s_{\rm vac}-1}.
\end{equation}
Assigning, as above, a duration $\delta t$ to each step, and rescaling time with the mean lifespan $\mu^{-1}$, the probability distribution for the duration of the vacant period, $T_{\rm vac}=\mu s_{\rm vac} \delta t$, reads
\begin{equation} \label{Pvac}
    P_T(T_{\rm vac}) = \frac{\alpha N}{\mu} \exp \left(-\frac{\alpha N}{\mu} T_{\rm vac}\right),
\end{equation}
in the limit $\delta t\to 0$. Here, $\alpha$ is defined as in Eq.~(\ref{alphanu}). Thus, the duration of vacant periods is exponentially distributed, with mean value $\langle T_{\rm vac} \rangle = \mu/\alpha N$. In the upper panel of Fig.~\ref{fig05}, we show normalized histograms for $T_{\rm vac}$. Each of them has been constructed by measuring the duration of $10^4$ vacant periods along numerical realizations of the model, for the same parameters as in Fig.~\ref{fig01} and four values of $N$. Straight lines correspond to the theoretical prediction of Eq.~(\ref{Pvac}), showing an excellent agreement with numerical results.

On the other hand, for small $\alpha$, the duration of  each active period, $T_{\rm act}$, is mainly controlled by the interplay of birth and death events. In particular, birth is a nonlinear process requiring the simultaneous presence of occupied and empty sites --cf.~$f_+ (n)$ in Eq.~(\ref{fs})--  which hinders an exact analytical treatment of its interaction with death, although large-$N$ asymptotic results for the mean value of  $T_{\rm act}$ have been reported \cite{Doering}. To gain insight into the statistics of active periods, we have run numerical realizations with initial population size $n_0=1$, up to the disappearance of the population, $n=0$, thus simulating a single active period. In these realizations, we have fixed $u=0.5$ and $p_\mu=0.2$, as in those shown in Fig.~\ref{fig01}. Moreover, since immigration is now not needed to restart the population activity, we have taken $p_\alpha=0$. For each run, we have recorded the duration $T_{\rm act}$, the time-averaged population density $\bar \rho_{\rm act}$, and the largest density reached along time, $\rho_{\rm max}$. The lower panel of Fig.~\ref{fig05} shows histograms for the active times $T_{\rm act}$, constructed from $10^4$ runs for each of four values of $N$ and $\nu/\mu=2.5$. The most noticeable feature in these results is the bimodal distribution of $T_{\rm act}$, which becomes better defined as $N$ grows. While for $N=5$ all runs aggregate into a single group with maximum frequency at $T_{\rm act}\approx 10$, for $N=30$ they have become segregated into two widely separated groups, with typical active times $T_{\rm act} \lesssim 1$ and $T_{\rm act}\gtrsim 10^4$, respectively.

\begin{figure}[ht]
\includegraphics[width=\columnwidth]{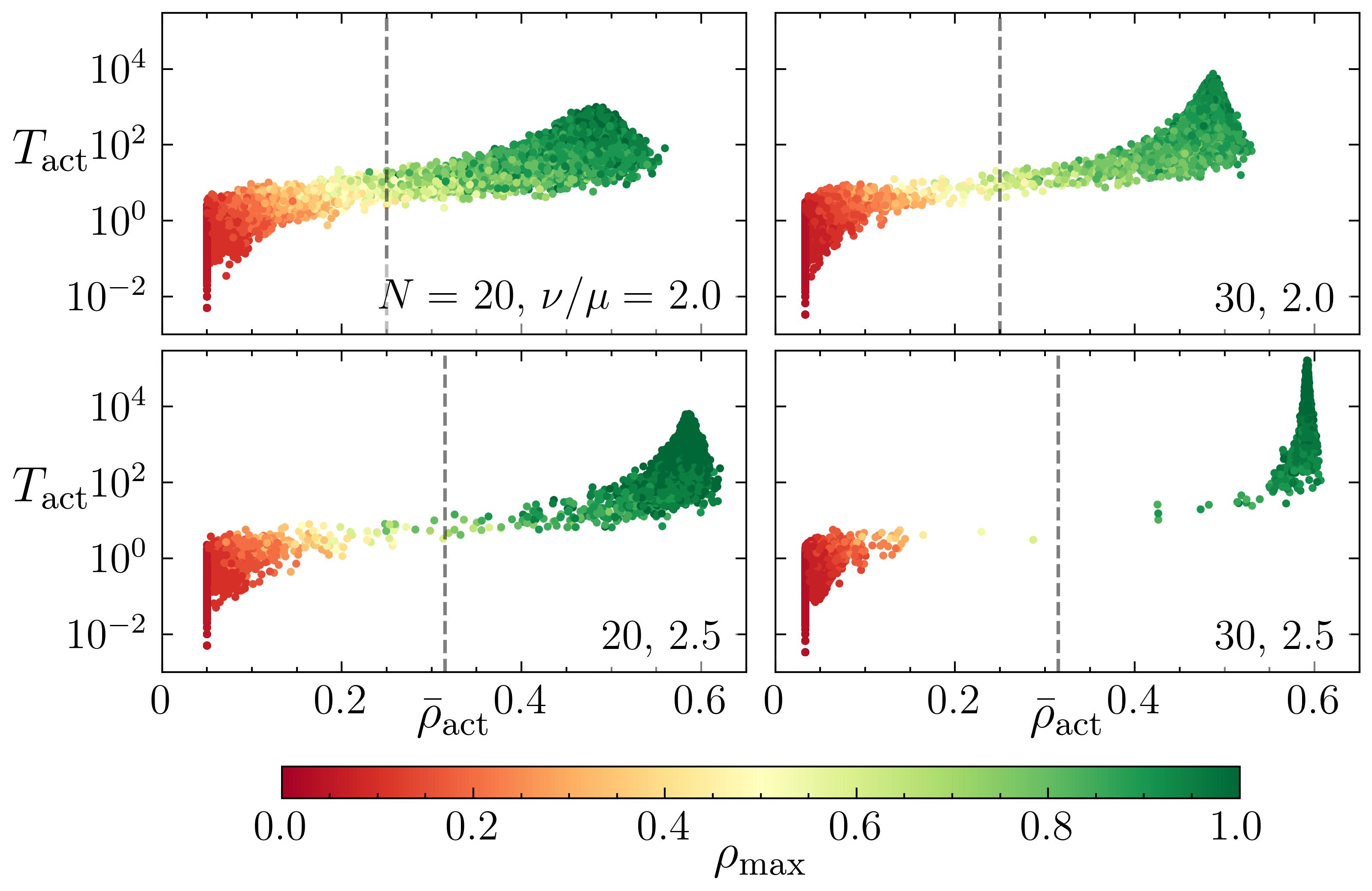}
\caption{\label{fig06}Scatter plots of active periods in the plane spanned by the time-averaged density $\bar \rho_{\rm act}$ and the duration $T_{\rm act}$ of each period, for four combinations of $N$ and $\nu/\mu$. Other parameters are as in Fig.~\ref{fig01}. Each dot corresponds to a single active period, with $10^4$ realizations in each panel.  Color indicates the largest density attained during the active period, $\rho_{\rm max}$, according to the scale at the bottom. Vertical dashed lines separate the groups of short and long active periods, following the criterion explained in the text.}
\end{figure}

This bimodality also reflects in the distribution of the time-averaged and largest densities along active periods, $\bar \rho_{\rm act}$  and  $\rho_{\rm max}$, as illustrated in Fig.~\ref{fig06} for four combinations of $N$ and $\nu/\mu$. In this figure, each dot represents a single run, out of a set of $10^4$ runs for each combination, and its coordinates stand for the pairs $(\bar \rho_{\rm act}, T_{\rm act})$ obtained for that run. Color, meanwhile, indicates the corresponding value of $\rho_{\rm max}$. The bimodal distribution of pairs $(\bar \rho_{\rm act}, T_{\rm act})$, with two abundant groups at respectively small and large values of both coordinates that become better separated as both $N$ and $\nu/\mu$ grow, is apparent. For $N=30$ and $\nu/\mu=2.5$ (lower-right panel) there are two well-defined classes of runs: short active periods where the density attains values of a few times the minimum $N^{-1}$, and long active periods --lasting at least four orders of magnitude longer than the former-- with average densities around $0.6$ and largest densities at, or close to, full occupation ($n=N$). Between the two groups, durations reach differences by a factor of $10^7$.

To analyze the statistical properties of the two groups, we must first classify the runs accordingly, as follows. For each parameter set, we rank the runs in order of increasing $\bar \rho_{\rm act}$. A plot of $\bar \rho_{\rm act}$ versus rank exhibits a clear inflection point, marking the transition between the two groups. We locate this inflection point by fitting a sigmoidal profile, given by a rational combination of exponential functions, which yields the value of $\bar \rho_{\rm act}$ defining the boundary between groups. This boundary is shown as a vertical dashed line in each panel of Fig.~\ref{fig06}.

\begin{figure}[ht]
\includegraphics[width=.8\columnwidth]{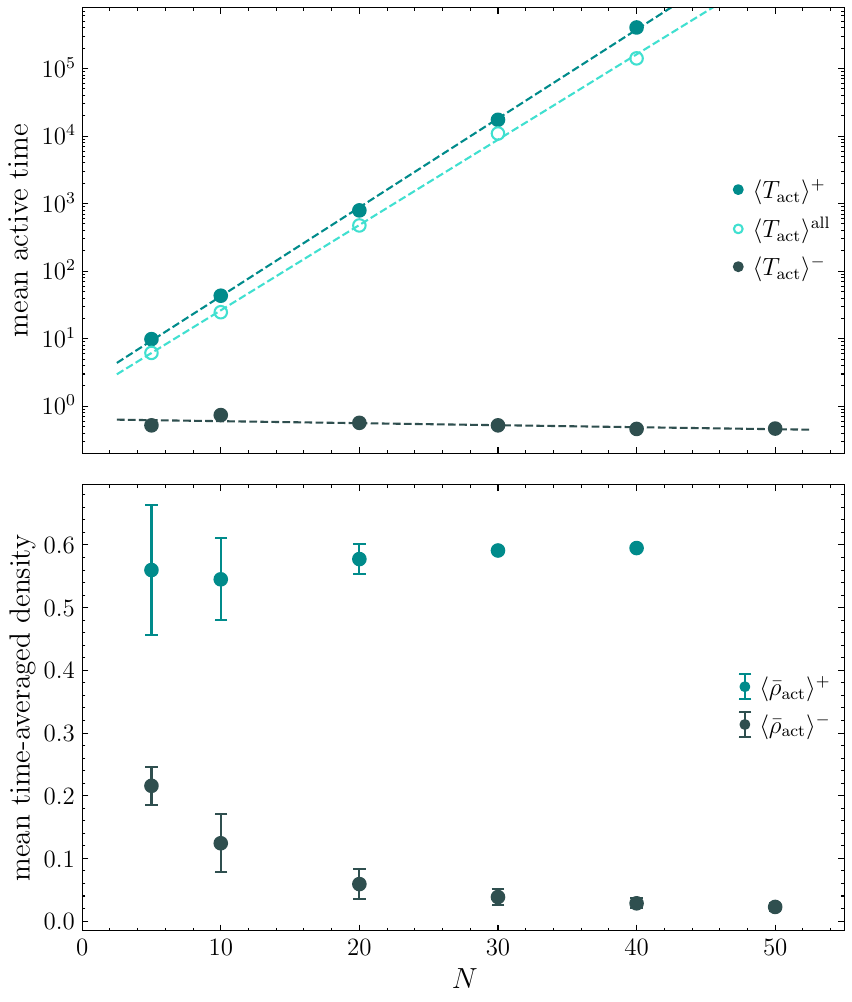}
\caption{\label{fig07}Upper panel: Full dots represent the mean duration of active periods, averaged separately over short and long runs, namely $\langle T_{\rm act} \rangle^-$ and $\langle T_{\rm act} \rangle^+$, for different values of $N$. The parameters are the same as in the lower panels of Fig.~\ref{fig06} ($\nu/\mu=2.5$). Empty dots indicate the corresponding average computed over the entire set of $10^4$ runs, $\langle T_{\rm act} \rangle^{\rm all}$. Straight dashed lines show linear fits for each dataset, highlighting the exponential growth of $\langle T_{\rm act} \rangle^+$ and $\langle T_{\rm act} \rangle^{\rm all}$, as well as the slight decrease of $\langle T_{\rm act} \rangle^-$.
Lower panel: As in the upper panel, for the mean time-averaged densities, $\langle \bar \rho_{\rm act} \rangle^{\pm}$, computed over the same realizations. Error bars represent the mean square dispersion within each set.
} 
\end{figure}

Figure \ref{fig07} displays statistical features measured for each group as a function of the maximum population size $N$. In the upper panel, full dots represent the mean active times for short and long runs, $\langle T_{\rm act} \rangle^-$ and $\langle T_{\rm act} \rangle^+$, respectively. In this linear-log plot, the straight dashed lines indicate an exponential dependence on $N$. Whereas $\langle T_{\rm act} \rangle^-$ decreases slowly, $\langle T_{\rm act} \rangle^+$ exhibits a rapid exponential growth with $N$. The lower panel presents the mean time-averaged density for each group, $\langle \bar \rho_{\rm act} \rangle^-$ and $\langle \bar \rho_{\rm act} \rangle^+$. As $N$ increases, both quantities quickly converge toward well-defined constant values.

In the upper panel of Fig.~\ref{fig07} we also plot, with open dots, the mean active time computed over the full set of $10^4$ runs, $\langle T_{\rm act} \rangle^{\rm all}$. Its exponential growth confirms the earlier result of Doering et al.~\cite{Doering}, who predicted this behavior in the large-$N$ limit when the birth rate exceeds mortality ($\nu > \mu$). Our results indicate that the exponential increase of $\langle T_{\rm act} \rangle^{\rm all}$ is primarily controlled by the group with long active times, $\langle T_{\rm act} \rangle^+$. Most likely, the transition from a unimodal to a bimodal distribution of active times --a feature not identified by those authors-- accounts for the change in the dependence of $\langle T_{\rm act} \rangle^{\rm all}$ on $N$ that they reported for the case without immigration, from a logarithmic dependence for $\nu<\mu$ to an exponential growth for $\nu>\mu$. 

The occurrence of two kinds of active periods, respectively with short and long durations, can be understood by assimilating the dynamics of the population size $n(t)$ to a one-dimensional random walk subjected to a force field, with $n(t)$ playing the role of the walker's position. In this representation, $n=0$ acts as an absorbing boundary condition where each realization of the random walk --and, thus, of the respective active period-- terminates. In order to analytically describe the random walk, it is convenient to derive the Fokker-Planck equation corresponding to the master equation (\ref{ME}). This amounts to approximating $n$ as a continuous variable, and expanding $P_{n\pm 1} =P_n \pm \partial_nP_n + \frac{1}{2} \partial^2_n P_n \pm \dots$ up to the appropriate order. The result is
\begin{equation}
    \partial_t P+ \partial_n \left( F_n P_n \right) =\partial_n \left( D_n \partial_n P_n \right),
\end{equation}
with a force 
\begin{eqnarray}
    F_n= && -\frac{\nu n^2}{N-1} +\left(\nu \frac{N+1}{N-1}-\mu \right)n \nonumber \\ &&-\frac{1}{2} \left( \nu\frac{N}{N-1}+\mu \right),
\end{eqnarray}    
and with diffusivity
\begin{equation}
    D_n= \frac{1}{2} \left( \nu \frac{N+1}{N-1}+\mu \right)n
\end{equation}
for $\alpha=0$. The force $F_n=-\partial_n U_n$ derives from a potential given by
\begin{eqnarray} \label{Un}
    U_n= &&\frac{\nu n^3}{3(N-1)}-\frac{1}{2}\left(\nu \frac{N+1}{N-1}-\mu \right)n^2 \nonumber \\ && +\frac{1}{2} \left( \nu\frac{N}{N-1}+\mu \right) n,
\end{eqnarray}
up to an arbitrary additive constant. 

\begin{figure}[ht]
\includegraphics[width=.8\columnwidth]{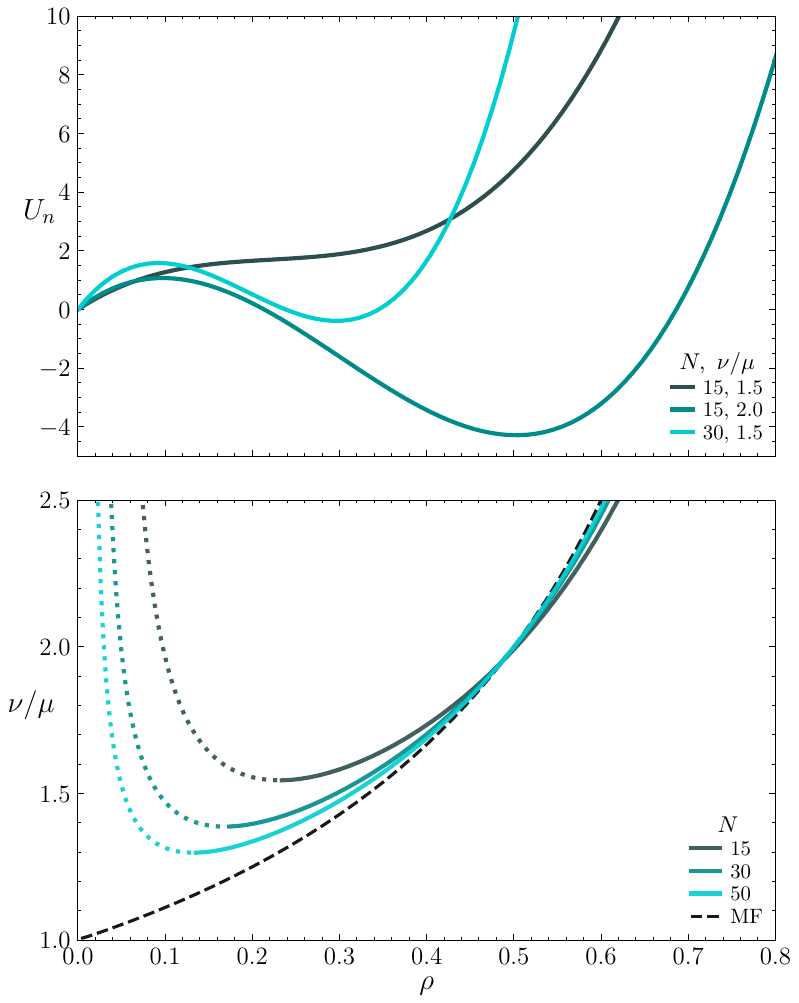}
\caption{\label{fig08} Upper panel: The potential $U_n$, Eq.~(\ref{Un}), as a function of $\rho=n/N$, for three combinations of $N$ and $\nu/\mu$. Lower panel: Positions of the minimum (full curve) and the maximum (dotted curve) of the potential $U_n$, as functions of $\nu/\mu$, for three values of $N$. The dashed curve represents the mean-field result,  $\rho^*_{\rm MF}$, as given by Eq.~(\ref{rhoMF}) with $\alpha=0$. For comparison with the upper panel, the horizontal and vertical axes have been exchanged with each other.} 
\end{figure}

The upper panel of Fig.~\ref{fig08} shows the plot of the potential $U_n$ as a function of $\rho=n/N$ for three combinations of $N$ and $\nu/\mu$. While for small values of these two parameters, $U_n$ is a monotonously increasing function, for sufficiently large values of $N$ and  $\nu/\mu$, the potential exhibits a maximum and a minimum for positive $\rho$. Specifically, these two extrema exist if the inequality 
\begin{equation} \label{cond}
    \left(N+1-\frac{N-1}{\nu/\mu} \right)^2> 2\left(  N+\frac{N-1}{\nu/\mu}\right)
\end{equation}
holds. The lower panel of the same figure illustrates how the position of the maximum (dotted curve) and the minimum (full curve) vary with $\nu/\mu$ for three values of $N$. Note that, for large $N$, the minimum approaches the mean field result $\rho^*_{\rm MF}$ given by Eq.~(\ref{rhoMF}) with $\alpha=0$ (dashed curve).

When the condition of Eq.~(\ref{cond}) holds, a random walker starting near $n=0$ and evolving under the potential $U_n$ may follow two qualitatively different paths, depending solely on stochastic fluctuations. If the walker remains to the left of the maximum, it has a high probability of reaching the absorbing boundary at $n=0$, terminating the walk after only a few steps. Conversely, if the walker overcomes the maximum, it will likely remain trapped for a long time in the potential well, producing a much longer walk. It may even cross the maximum multiple times before eventually being absorbed at $n = 0$. This bimodality is reflected in the distributions of durations and mean densities during the active periods of our model, as illustrated by Figs.~\ref{fig05} and \ref{fig06}. 

\section{Stochastic logistic dynamics on networks}

In order to consider distributed systems, we extend our model assuming that the $N$ sites available to the population are situated at the nodes of a network with controlled structural properties. As detailed below, we consider three classes of random networks with different architectures: Erd\H{o}s-R\'enyi, regular \cite{Newman}, and geographic \cite{Masuda} networks, the latter with periodic boundary conditions. On networks, the dynamical rules for death and immigration events are exactly the same as for unstructured populations (see Section \ref{II}). In turn, the ordered pair of sites selected for a birth event must consist of two sites connected by a network link. Consequently, whenever a birth occurs, the parent and offspring are neighbors.

The constraint that birth events must occur at neighboring sites is expected to localize the population into one or more aggregates occupying mutually connected network nodes. This aggregation, in turn, should retard population growth owing to the reduced availability of free sites for placing newborns. Consequently, in the long run, the average population size should be smaller than when births are not restricted by the network structure. At the same time, longer vacant periods are expected to occur. In Sections \ref{nets} and \ref{asso}, we characterize growth retardation and population aggregation by comparing vacant times and average activity with the case of unstructured populations, and by studying occupancy assortativity, respectively.

Our analysis of the population dynamics on networks is performed numerically. In each realization, we first build an $N$-node random network of a given class, with a total of $M$ links between nodes. For {\em Erd\H{o}s-R\'enyi} networks, we distribute the $M$ links at random over the set of nodes. The resulting average number of links per node is $\langle k\rangle =2M/N$. For {\em regular} networks, $k=2M/N$ links are assigned to each node and used to connect it with exactly $k$ nodes chosen at random over the network (naturally, for $k$ to be an integer number, $M$ must be an integer multiple of $N/2$).  In both cases,  self-connections and multiple connections between any two nodes are avoided. Finally, {\em geographic} networks are built by, first, uniformly distributing the $N$ nodes at random over a planar square domain. Then, the $M$ pairs of nodes with the shortest mutual Euclidean distances are joined by a single link. To avoid an uneven distribution of links between nodes located near the center and those near the edges of the square, the construction is implemented using periodic boundary conditions. This is equivalent to extending the system over the entire plane by replicating the original square domain in both directions. For the three classes of networks, we discard realizations in which the resulting network is not connected, namely, when it consists of more than one connected component. This condition prevents the population from being confined to a subnetwork of size smaller than $N$.

\begin{figure}[ht]
\includegraphics[width=.8\columnwidth]{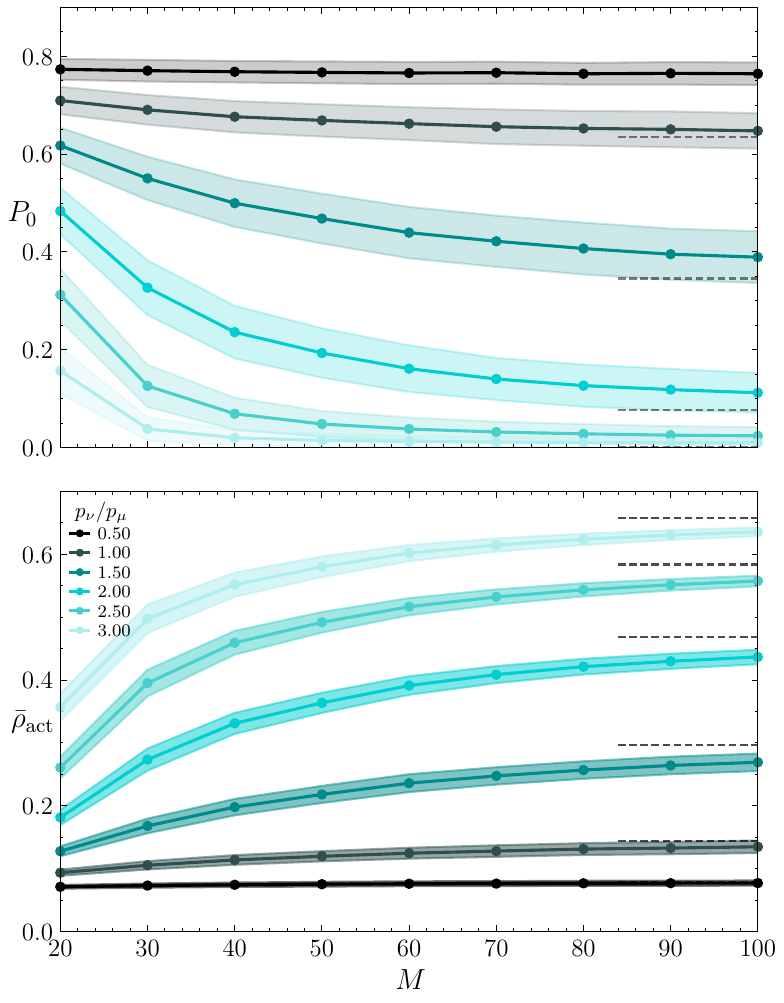}
\caption{\label{fig09} Upper panel: Fraction of vacant time, $P_0$, measured in populations evolving on Erd\H{o}s-R\'enyi networks, for $N=20$, $u=0.5$, $p_\mu=0.2$,  $p_\alpha/p_\mu=0.01$ and several values of $p_\nu/p_\mu$, as indicated in the labels of the lower panel. Each dot corresponds to the average over $2500$ realizations. For clarity, dots have been joined by straight segments. The colored bands represent the mean square dispersion over realizations. The horizontal dashed segments to the right stand for the values of $P_0^*$, derived from the master equation (\ref{ME}). Lower panel: As in the upper panel, for the time-averaged density in active periods, $\bar \rho_{\rm act}$.} 
\end{figure}

Figure \ref{fig09} illustrates numerical results for the case of Erd\H{o}s-R\'enyi networks. The upper panel shows the measured fraction of vacant time, $P_0$, as a function of the number of links in the network, $M$, for $N=20$, $u=0.5$, $p_\mu=0.2$, $p_\alpha/p_\mu=0.01$, and several values of $p_\nu/p_\mu$.  Colored bands represent the mean square dispersion over sets of $2500$ realizations. Each realization was run on a different network. We see that the largest values of $P_0$ are obtained for small $M$, where network links are more sparse and the effect of the underlying structure is thus larger. As $M$ grows and the network becomes better connected, $P_0$ approaches the values predicted for the unstructured system, $P_0^*$, as obtained from the stationary solution of the master equation (\ref{ME}). Concurrently, as expected, $P_0$ decreases as the birth rate grows. Meanwhile, the time-averaged density in active periods $\bar \rho_{\rm act}$, shown in the lower panel, exhibits the opposite behavior, increasing with both $M$ and $p_\nu/p_\mu$. Although with minor quantitative differences, results for regular and geographic networks show the same qualitative trends.

\subsection{Vacant time and average activity:  Unstructured vs. distributed systems} \label{nets}

For a more thorough comparison of the quantities studied in Sections \ref{IIB} and \ref{IIC} between unstructured and distributed populations, we first consider the ratio $\bar P_0 /P_0^*$, between the fraction of time spent in vacant periods on networks, averaged over sets of $2500$ realizations, $\bar P_0$, and the corresponding quantity for unstructured populations, $P_0^*$. Results for different network architectures are compared by fixing the number of links $M$. Generally, we expect that quantitative results for distributed populations approach those of unstructured systems when $M$ increases.  

\begin{figure}[ht]
\includegraphics[width=.75\columnwidth]{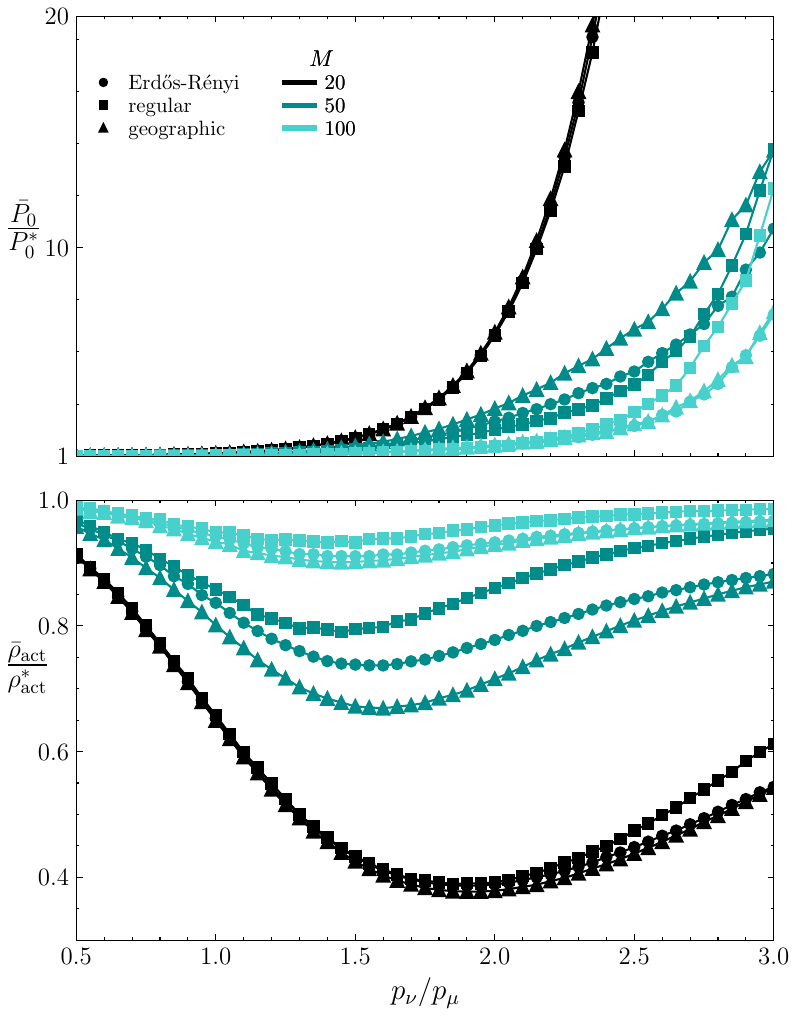}
\caption{\label{fig10} Upper panel: Ratio of the fractions of vacant time measured in populations evolving on networks averaged over 2500 realizations, $\bar P_0$, and those derived from the master equation (\ref{ME}) for unstructured populations, $P_0^*$, as a function of $p_\nu/p_\mu$, for $N=20$, $u=0.5$, $p_\mu=0.2$, and $p_\alpha/p_\mu=0.01$. As indicated in the labels, results correspond to three network architectures and three values of the number of network links, $M$. Lower panel: As in the upper panel, for the ratio of average densities along active periods.} 
\end{figure}

The upper panel of Fig.~\ref{fig10} shows numerical results for the ratio $\bar P_0 /P_0^*$ as a function of $p_\nu/p_\mu=\nu/\mu$, for systems of size $N=20$, with $u=0.5$, $p_\mu=0.2$, and $p_\alpha/p_\mu=0.01$. Different sets, joined by lines for clarity, correspond to the three network architectures and three values of $M$, as indicated in the legend. We find that, in all cases, $\bar P_0 /P_0^*>1$.  In other words, compared with its unstructured counterpart, a population evolving on a network always spends a larger fraction of time in vacant periods. The distribution of vacant durations, however, does not depend on the underlying structure of the system, since the termination of vacant periods is determined exclusively by immigration; cf. Eq.~(\ref{Pvac}). Therefore, the fact that  $\bar P_0 /P_0^*>1$  implies that, on networks, the {\em number} of vacant periods is systematically larger than for unstructured populations. That is to say, events in which the population becomes extinct are more frequent in the former than in the latter. 

As for the dependence on the birth rate, the ratio $\bar P_0 /P_0^*$ shows a fast growth with $p_\nu/p_\mu$, starting from values near unity at small birth rates. As expected, this effect is more pronounced for small $M$.  Comparing between network architectures, results for the three classes exhibit the same qualitative behavior, with moderate quantitative differences as $M$ changes.  

For the same numerical realizations on networks, we have computed the ratio between the time-averaged density during active periods and the same quantity in unstructured systems, $\bar \rho_{\rm act} /\rho^*_{\rm act}$. Results are shown in the lower panel of Fig.~\ref{fig10}. In all cases, $\bar \rho_{\rm act} /\rho^*_{\rm act} < 1$, indicating that average population sizes on networks are systematically lower than in the unstructured case. This is directly consistent with the conjecture that spatial structure limits population growth, due to the aggregation of occupied sites into localized clusters. The same picture explains the most conspicuous feature of $\bar \rho_{\rm act}/ \rho^*_{\rm act}$, namely, its non-monotonic dependence on $p_\nu/p_\mu$. In fact, the effect of population clustering is expected to be larger for moderate densities, when the number of occupied (or empty) sites is neither too small nor too close to $N$. These two extremes correspond to the limits of small and large birth rates where, as seen in Fig.~\ref{fig10}, the ratio $\bar \rho_{\rm act}/ \rho^*_{\rm act}$ is indeed closer to unity.       

Much as seen in the upper panel for the ratio of vacant times,  $\bar \rho_{\rm act}/ \rho^*_{\rm act}$ exhibits the same qualitative dependence on $p_\nu/p_\mu$ for all the parameter sets considered in the numerical realizations. As above, results on networks approach those for unstructured populations when the average number of links $M$ grows. Different network architectures do not affect the overall behavior, but induce rather disparate values of $\bar \rho_{\rm act}/ \rho^*_{\rm act}$ for intermediate values of $p_\nu/p_\mu$ and $M$. Note that, in all cases, the differences with the unstructured case are larger for geographic networks. This is likely a direct consequence of the fact that, when nodes are effectively distributed in space and connected only to nearby neighbors, the clustering of occupied nodes is enhanced.

\subsection{Occupancy assortativity} \label{asso} 

In order to quantify the propensity of the population to form clusters when evolving on networks, we study the occupancy assortativity, which measures the tendency of occupied sites to be neighbors of each other. We recall that, for a scalar attribute $z_i$ assigned to each node $i$, the assortativity $a_z$ is defined as the Pearson correlation coefficient of the quantity $z$ over the set of connected pairs of nodes. For a network with $M$ links, it can be computed as  \cite{Newman}
\begin{equation} \label{az}
    a_z= \frac{\sum_{i,j} (A_{ij} -k_ik_j/2M) z_iz_j}{\sum_{i,j} (k_i\delta_{ij} -k_ik_j/2M) z_iz_j}, 
\end{equation}
where the sums run over all the pairs of nodes, $k_i$ is the degree (number of neighbors) of node $i$, $A_{ij}$ is the adjacency matrix of the network ($A_{ij}=1$ if nodes $i$ and $j$ are neighbors, and $0$ otherwise), and $\delta_{ij}$ is the Kronecker delta. By construction, $-1\le a_z\le 1$. Positive (respectively, negative) assortativity is obtained when the values of $z$ in neighbor nodes are mutually similar (respectively, disparate). In our case, we determine the assortativity for the occupancy function defined as
\begin{equation}
    z_i=\begin{cases}
        1 & \mbox{if site $i$ is occupied,} \\
        0  & \mbox{otherwise}.
    \end{cases}
\end{equation}

The occupancy assortativity is expected to depend not only on the parameters governing the dynamics and the network architecture, but also on the instantaneous population size. Indeed, when the number of occupied sites $n$ is very small or very close to the maximum $N$, a clustered distribution cannot be statistically distinguished from a random one. Consequently, significant values of $a_z$ are expected for intermediate values of $n$. To compare the occupancy assortativity with a reference level, we first consider its value for a random distribution of occupied sites, which is expected to provide a lower bound for the actual values of $a_z$. Averaging over random distributions in Eq.~(\ref{az}), we get
\begin{equation} \label{azrand}
    a_z^{\rm rand} = - \frac{1}{N-1},
\end{equation}
for $n\neq 0$ and $n\neq N$, which does not depend on $n$. For $n=0$ and $n=N$, the assortativity is ill-defined. Second, to obtain an upper bound for $a_z$, we numerically construct ``artificial'' distributions of occupied sites on networks of each architecture by maximizing the occupancy in the neighborhood of each occupied site, and measuring the resulting assortativity as a function of the population size $n$. Its average over many realizations of the network structure, $a_z^{\rm max}$,  is expected to give a suitable estimation of the largest value of $a_z$ for each $n$.

\begin{figure}[ht]
\includegraphics[width=.75\columnwidth]{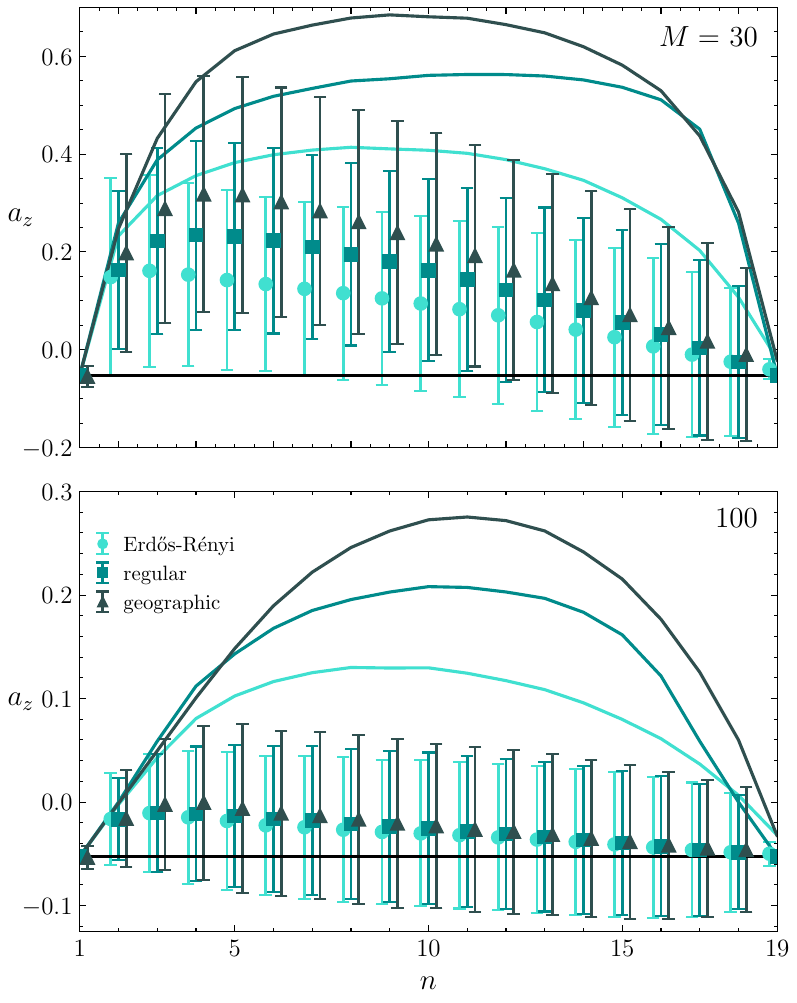}
\caption{\label{fig11} Occupancy assortativity $a_z$ calculated as in Eq.~(\ref{az}) as a function of the population size $n$, in sets of numerical realizations of the system for $N=20$, $u=0.5$, $p_\alpha=0.002$, $p_\mu=0.2$, and $p_\nu/p_\mu=2$, over three networks classes, with $M=30$ (upper panel) and $100$ (lower panel). Symbols stand for averages of $a_z$ over realizations, and error bars indicate the corresponding standard deviations. For clarity, data for different network classes have been slightly shifted along the horizontal direction. The horizontal line indicates the assortativity of random distributions, $a_z^{\rm rand}$, as given by Eq.~(\ref{azrand}), and curves are the upper bounds $a_z^{\rm max}$ for each network class, determined as explained in the text.}
\end{figure}

We have measured the occupancy assortativity in numerical realizations of our system for $N=20$, $u=0.5$, $p_\mu=0.2$, and $p_\alpha/p_\mu=0.01$, over the three network classes. In each realization, after a certain transient time, the population size $n$ was recorded and the instantaneous value of $a_z$ was calculated using Eq.~(\ref{az}). We have found that, remarkably, $a_z$ is virtually independent of the birth rate, at least, in the investigated interval: $1\lesssim p_\nu/p_\mu \lesssim 3$. Symbols in Fig.~\ref{fig11} stand for the occupancy assortativity $a_z$ averaged over realizations as a function of $n$. Error bars indicate the corresponding standard deviations. The horizontal line indicates the expected lower bound given by the assortativity of random distributions, Eq.~(\ref{azrand}), and the curves are the upper bounds $a_z^{\rm max}$ for each network architecture, determined as explained above. The upper and lower panels of the figure respectively correspond to networks with $M=30$ and $100$ links. Although  $a_z$ exhibits considerable dispersion over realizations, its average does reach significant values for intermediate population sizes $n$. As expected, this average coincides with $a_z^{\rm rand}$ and $a_z^{\rm max}$ when $n=1$ and $N-1$, where the distribution of occupied sites cannot be discerned from a random distribution. It attains a maximum for relatively low values of $n$, and decays slowly as $n$ grows further. As with the other quantities examined so far, the influence of network structure on assortativity becomes attenuated as $M$ increases, with $a_z$ approaching the values expected for random distributions. For the two values of $M$ shown in the figure, the deviation from the unstructured case is larger for geographic networks, in agreement with the results obtained above for the population density during active periods.

The upper half of the animation included as Supplemental Material \cite{SM} compares a population with $N=50$ evolving on a geographic network (left) and in an unstructured system (right). To ease the comparison, sites in the unstructured system are plotted in the same positions as for the geographic network although, in the former, their spatial distribution is irrelevant to the dynamics. Dots outside the square in the leftmost plot represent the sites connected to sites on the opposite side of the square, due to the periodic boundary conditions. Big dark dots indicate occupied sites at each time. In contrast with the unstructured system, the progressive formation of aggregates on the geographic network is apparent. In the lower plot of the animation, we show the time dependence of the instantaneous occupancy assortativity $a_z$ for both systems. As expected, $a_z$ is predominantly higher for the geographic network.      
\section{Conclusion}

In this paper we have analyzed a stochastic model for the dynamics of a population driven by birth, death, and immigration events. By construction, the model imposes a maximum population size, making it particularly suitable for studying fluctuations arising from the discreteness of the system. This intrinsic size limitation acts as an environmental carrying capacity \cite{Murray}, so that, in the mean-field limit and in the absence of immigration, the dynamics reduce to logistic evolution. To a large extent, our focus has been on the dynamical effects of discreteness in comparison with the mean-field behavior.

The evolution of our model consists of an alternation of active periods, where the population size is positive, and vacant periods, where the population has become extinct. Vacant periods are terminated by immigration events, thus initiating active periods. We have statistically characterized this behavior by analyzing the frequency with which the population reaches each possible size over time, a problem that can be efficiently addressed by solving a master equation in the stationary regime. In contrast, characterizing the distributions of durations and population sizes during individual active periods required extensive numerical realizations of the model. Our results show that, in sharp contrast with the mean-field behavior, fluctuations can induce two coexisting kinds of activity: short active periods, where the population size typically remains around a small fraction of its maximum, and long active periods, where the population size sustainedly fluctuates around large values, and can even attain the maximum. In turn, the durations of short and long active periods can differ by several orders of magnitude.    

Analytical and numerical stochastic approaches to the study of the duration of active periods --also referred to in the literature as extinction or persistence times-- have been developed for several ecological systems over at least the past two decades. We have already mentioned the work by Doering et al.~\cite{Doering}, in which the authors derived large-size asymptotic results for the mean extinction time of a stochastic logistic population without immigration. Our results are compatible with their predictions. Explicit functional forms for the distribution of persistence times have also been obtained for systems of several competing species in bounded domains \cite{Pigolotti,Suweis}, and a comparison with real ecosystems has also been attempted \cite{Bertuzzo}. In this case of inter-species competition, the results typically display power-law distributions with exponential cutoffs. The marked contrast with the bimodal distributions reported here for intra-species logistic competition suggests that persistence times are highly sensitive to the particular ecological setting being considered.

While the results summarized above pertain to spatially unstructured systems, where all sites available to the population are mutually neighboring, we have also analyzed a variant of the model in which the population is distributed over a network. This distribution primarily affects birth events, as reproduction requires that the sites occupied by parents and offspring be mutual neighbors. The main consequence of this constraint is that, at long times, the population tends to aggregate into relatively small clusters of neighboring sites. This clustering, in turn, slows the dynamics, since only individuals located at the borders of clusters --those with empty neighboring sites-- are able to reproduce. The overall effect is a reduction in the average population size relative to unstructured systems. We have characterized this phenomenon, demonstrating not only a decrease in population size but also an increased frequency of vacant periods. We have further quantified the degree of clustering by measuring the assortativity of occupancy as a function of population size. This analysis was carried out on three types of random networks. Although the qualitative consequences of an underlying network structure were similar across the three types, the effect was stronger in geographic networks, where the structure is directly associated with an actual spatial distribution. As expected, results for populations on networks approach those of unstructured systems when the network connectivity grows.  

Several extensions can be envisaged to bring the present model closer to more realistic situations. First, in a small population, sexual reproduction plays a crucial limiting role, since individuals of both sexes are required for the population to grow by birth. Another limiting factor for population growth is age, as individuals must reach sexual maturity to be capable of reproduction. Introducing sex and age structure as individual attributes is therefore an interesting prospect for extending our results. These two attributes contribute to diversity at the individual level, which could be further enhanced by heterogeneity in the parameters controlling birth and death. Finally, in systems evolving on networks, heterogeneity may also be associated with attributes assigned to each network site, representing environmental diversity.

\bibliography{refs} 
\end{document}